\documentclass[final]{elsart3p}

\journal{J. Crystal Growth, for CGCT4 Sendai}
\date{22 May 2008}

\usepackage{graphicx}
\usepackage{amssymb}
\usepackage{color}

\begin{document}


\begin{frontmatter}

\title{Growth of Oxide Compounds under Dynamic Atmosphere Composition}

\author{D. Klimm\corauthref{cor1}}\ead{klimm@ikz-berlin.de},
\author{S. Ganschow},
\author{D. Schulz},
\author{R. Bertram},
\author{R. Uecker},
\author{P. Reiche}, and
\author{R. Fornari}

\corauth[cor1]{corresponding author}

\address{Leibniz Institute for Crystal Growth, Max-Born-Str. 2, 12489 Berlin, Germany}

\begin{abstract}
Commercially available gases contain residual impurities leading to a background oxygen partial pressure of typically several $10^{-6}$\,bar, independent of temperature. This oxygen partial pressure is inappropriate for the growth of some single crystals where the desired oxidation state possesses a narrow stability field. Equilibrium thermodynamic calculations allow the determination of dynamic atmosphere compositions yielding such self adjusting and temperature dependent oxygen partial pressures, that crystals like ZnO, Ga$_2$O$_3$, or Fe$_{1-x}$O can be grown from the melt.
\end{abstract}


\end{frontmatter}

\section{Introduction and thermodynamic background}

Crystal growth is often performed from melts contained in crucibles, e.g. by Czochralski or Bridgman technique. Chemical compatibility of the substance to be grown, the crucible material, and other constructional elements under the given temperature $T$ and atmosphere is an inevitable prerequisite for the establishment of stable growth conditions. In case of metal oxides MeO$_{m/2}$ ($m=0-8$ is the valency) usually crucibles made of another metal Me$^{'}$ with high melting point $T_\mathrm{f}$ are used: platinum ($T_\mathrm{f}=1770^{\,\circ}$C), iridium ($T_\mathrm{f}=2443^{\,\circ}$C), and tungsten ($T_\mathrm{f}=3407^{\,\circ}$C) are typical examples. (Thermodynamic data are taken from the FactSage databases \cite{FactSage5_5} throughout this article.) Unfortunately, the chemical stability of these refractory metals drops with rising $T_\mathrm{f}$.

The chemical stability of all components is governed by thermodynamic equilibria if the temperature is sufficiently high: this is usually true for $T\gg800-1000^{\,\circ}$C. In such cases the behavior of systems containing only pure condensed phases can be predicted easily considering redox reactions

\begin{equation}
2\; \mathrm{MeO}_{m/2} + \frac{1}{2}\; \mathrm{O}_2 \rightleftarrows 2\; \mathrm{MeO}_{(m+1)/2}
\label{eq:redox}
\end{equation}

for all metals Me (crucibles) or metal oxides MeO$_x$ (substances to be grown, insulating ceramics). The description of the system by (\ref{eq:redox}) for the Me$^{'}$, Me$^{''}$,\ldots\ is almost correct even if some of them are forming condensed phases with more then one metal (e.g. melts of several oxides, intermediate compounds MeMe$^{'}$O$_x$, or solid solutions (Me,Me$^{'}$)O$_x$), as the formation enthalpy of such phases from the MeO$_x$ is often small compared to $\mathit\Delta H$ for (\ref{eq:redox}).

At equilibrium $\mathit\Delta G$ for (\ref{eq:redox}) vanishes. If moreover the vapor pressures of the metals and their oxides can be neglected compared to the oxygen partial pressure $p_{\mathrm{O}_2}$, one finds for the equilibrium constant $K$ of reaction (\ref{eq:redox})

\begin{equation}
-RT \ln K = RT \ln p_{\mathrm{O}_2} = \mathit\Delta G^0 = \mathit\Delta H^0 - T \mathit\Delta S^0
\label{eq:Ellingham}
\end{equation}

where $\mathit\Delta G^0$ is the standard free energy change of (\ref{eq:redox}). As $\mathit\Delta H^0$ and $\mathit\Delta S^0$ change usually weakly with $T$ it is obvious from (\ref{eq:Ellingham}) that plots of $\mathit\Delta G^0 = RT \ln p_{\mathrm{O}_2}$ vs. $T$ (``Ellingham diagrams'') are straight lines for each oxidation reaction. These lines separate phase regions where one oxidation state prevails and the whole graph represents a predominance phase diagram for the corresponding metal Me and its oxides \cite{Pelton91,Klimm99c}. Ellingham and other predominance diagrams can be calculated either manually from tabulated $\mathit\Delta H^0$, $\mathit\Delta S^0$ \cite{Barin93} or with software packages like \mbox{FactSage} or \mbox{Thermocalc} \cite{FactSage5_5,Thermocalc08}, respectively, and are discussed in more detail elsewhere \cite{Pelton91,Atkins06,Klimm99b}.

\begin{figure}[htb]
\includegraphics[width=0.47\textwidth]{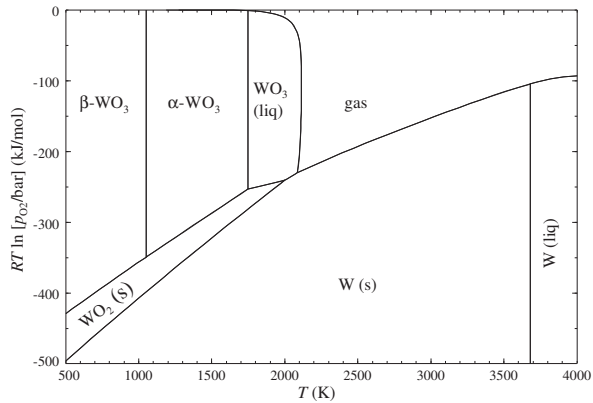}
\caption{Predominance diagram W--O$_2$ showing stability regions for tungsten (2 condensed phases), tungsten(IV) oxide (1 condensed phase), tungsten(VI) oxide (3 condensed phases), and gas (no condensed phases).}
\label{fig:w_o2}
\end{figure}

Fig.~\ref{fig:w_o2} shows an Ellingham diagram for the system tungsten--oxygen where transitions between different oxidation states (\ref{eq:redox}) are indicated by inclined lines and transformations between different pure phases (e.g. triclinic $\beta$-WO$_3$, tetragonal $\alpha$-WO$_3$ \cite{Kehl52}, and liquid WO$_3$) as vertical lines. It will be shown in the following that these diagrams are powerful tools for the manipulation of oxidation states during high $T$ crystal growth processes.

\section{Crystal growth of oxides}

\subsection{General considerations}

It is an inherent result of (\ref{eq:redox}) and (\ref{eq:Ellingham}) that $p_{\mathrm{O}_2}$ may never be zero to keep any metal oxide thermodynamically stable. If $p_{\mathrm{O}_2}$ is too low (\ref{eq:redox}) is shifted to the left hand side resulting in an oxide with lower oxidation state. For sufficiently low $p_{\mathrm{O}_2}$, finally the metal (valency $m=0$) is formed.

``Pure'' argon or nitrogen are often reported as ``inert'' gases to be used during crystal growth processes. ``Pure'' means usually 5N gases: Actually such gas represents a mixture of 99.999\% gas (e.g. Ar, N$_2$) + 0.001\% (10\,ppm) impurity. This impurity is not well defined, but could e.g. be air (ca. 21\% O$_2$ + ca. 79\% N$_2$). In this case, 5N nitrogen with total pressure $p=1$\,bar can be described as a constant admixture of $\approx2\times10^{-6}$\,bar O$_2$ in nitrogen. $p_{\mathrm{O}_2}=2\times10^{-6}$\,bar is too high for the growth of (La,Sr)(Al,Ta)O$_3$ (LSAT) crystals that are dark-red colored in the as-grown state and can be bleached by annealing in H$_2$ \cite{Sakowska01}. Other oxides are unstable at elevated $T$ under too low $p_{\mathrm{O}_2}$, and typically minor amounts of oxygen are added to the gas: La$_{1-x}$Nd$_x$GaO$_3$ from N$_2$ + 1\% O$_2$ \cite{Berkowski00}; Sm$_3$Ga$_5$O$_{12}$ or Gd$_3$Ga$_5$O$_{12}$ from N$_2$ with 0.9 or 1.5\% O$_2$, respectively \cite{Digiuseppe80}. The influence of oxygen partial pressures from $2\times10^{-5}$\,bar to $1\times10^{-2}$\,bar on the color of melt grown SrLaGaO$_4$ and SrLaAlO$_4$ was described by Pajaczkowska et al. \cite{Pajaczkowska99,Pajaczkowska01}.

\begin{figure}[htb]
\includegraphics[width=0.47\textwidth]{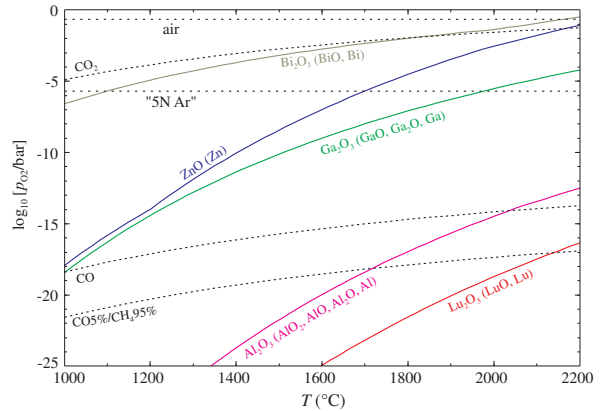}
\caption{Full lines: Minimum oxygen pressures $p_{\mathrm{O}_2}(T)$ that are necessary for the stabilization of several metal oxides (decomposition products for too low $p_{\mathrm{O}_2}$ given in brackets). Dashed lines: Equilibrium values $p_{\mathrm{O}_2}(T)$ that are created by several gases or gas mixtures ($p=1$\,bar).}
\label{fig:t_vs_ox}
\end{figure}

Obviously, by addition of O$_2$ to the growth atmosphere only a constant (independent of $T$) partial pressure $\approx 2\times10^{-6}\,\mathrm{bar} \lesssim p_{\mathrm{O}_2} \lesssim 1\,\mathrm{bar}$ can be maintained. In Fig.~\ref{fig:t_vs_ox} the level of the upper abscissa with $p_{\mathrm{O}_2}=\mathrm{const.}=1$\,bar represents pure oxygen and the horizontal dashed lines represent air ($p_{\mathrm{O}_2}=\mathrm{const.}=0.21$\,bar) or 5N argon with 10\,ppm air, respectively. As the reactions (\ref{eq:redox}) are mainly exothermal, the $p_{\mathrm{O}_2}$ that is required to stabilize a specific MeO$_x$ rises usually with $T$. Fig.~\ref{fig:t_vs_ox} demonstrates this for Lu$_2$O$_3$, Al$_2$O$_3$, Ga$_2$O$_3$, ZnO, and Bi$_2$O$_3$. These oxides are stable for all $p_{\mathrm{O}_2}(T)$ above the corresponding lines and tend to decompose to the species given in brackets for lower oxygen partial pressures. The stability of the oxides listed above requires growing $p_{\mathrm{O}_2}$.

The gaseous oxides CO$_2$, CO, and H$_2$O decompose with $T$ like the reverse reaction of (\ref{eq:redox})

\begin{eqnarray}
\mathrm{CO}_2 & \rightleftarrows & \mathrm{CO} + \half \, \mathrm{O}_2 \label{eq:CO2} \\
\mathrm{CO}   & \rightleftarrows & \mathrm{C} +  \half \, \mathrm{O}_2 \label{eq:CO} \\
\mathrm{H}_2\mathrm{O} & \rightleftarrows & \mathrm{H}_2 +  \half \, \mathrm{O}_2 \label{eq:H2O} \\
\mathrm{C} + 2\,\mathrm{H}_2 & \rightleftarrows & \mathrm{CH}_4 \label{eq:CH4}
\end{eqnarray}

and lead to the production of oxygen if $T$ rises. The reaction partners of (\ref{eq:CO2})--(\ref{eq:H2O}) and methane CH$_4$ are combined e.g. by (\ref{eq:CH4}). Fig.~\ref{fig:t_vs_ox} shows that mixtures of CO$_2$, CO, and CH$_4$ allow to adjust $p_{\mathrm{O}_2}$ over more than 10 orders of magnitude. Moreover, $p_{\mathrm{O}_2}(T)$ for such atmospheres changes similarly to many MeO$_x$. An appropriate gas composition can provide a ``self adjusting'' oxygen partial pressure that keeps the desired oxidation state thermodynamically stable over an extended $T$ range.

\subsection{Examples}

\begin{figure}[htb]
\includegraphics[width=0.42\textwidth]{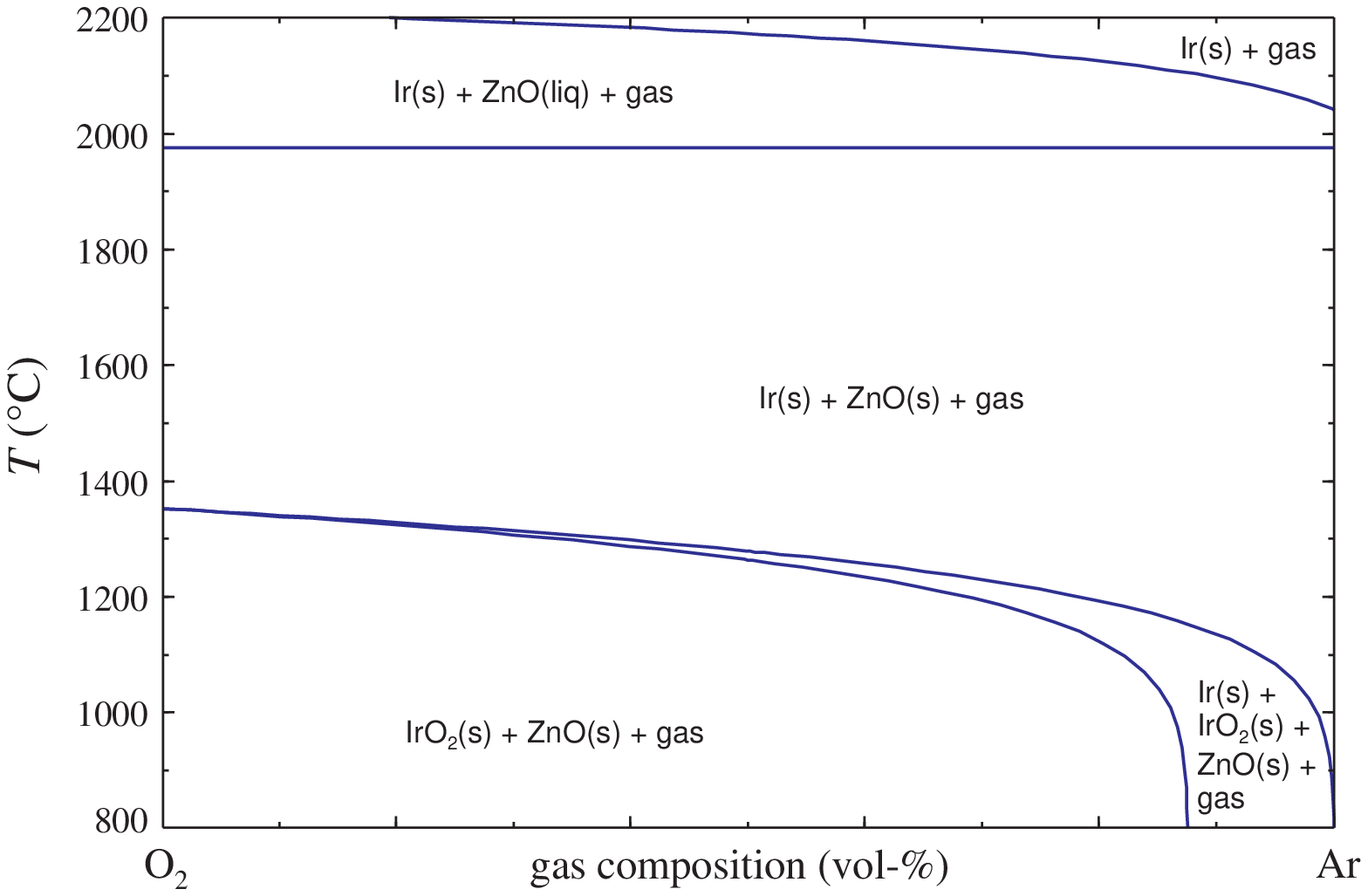} \ a)

\includegraphics[width=0.42\textwidth]{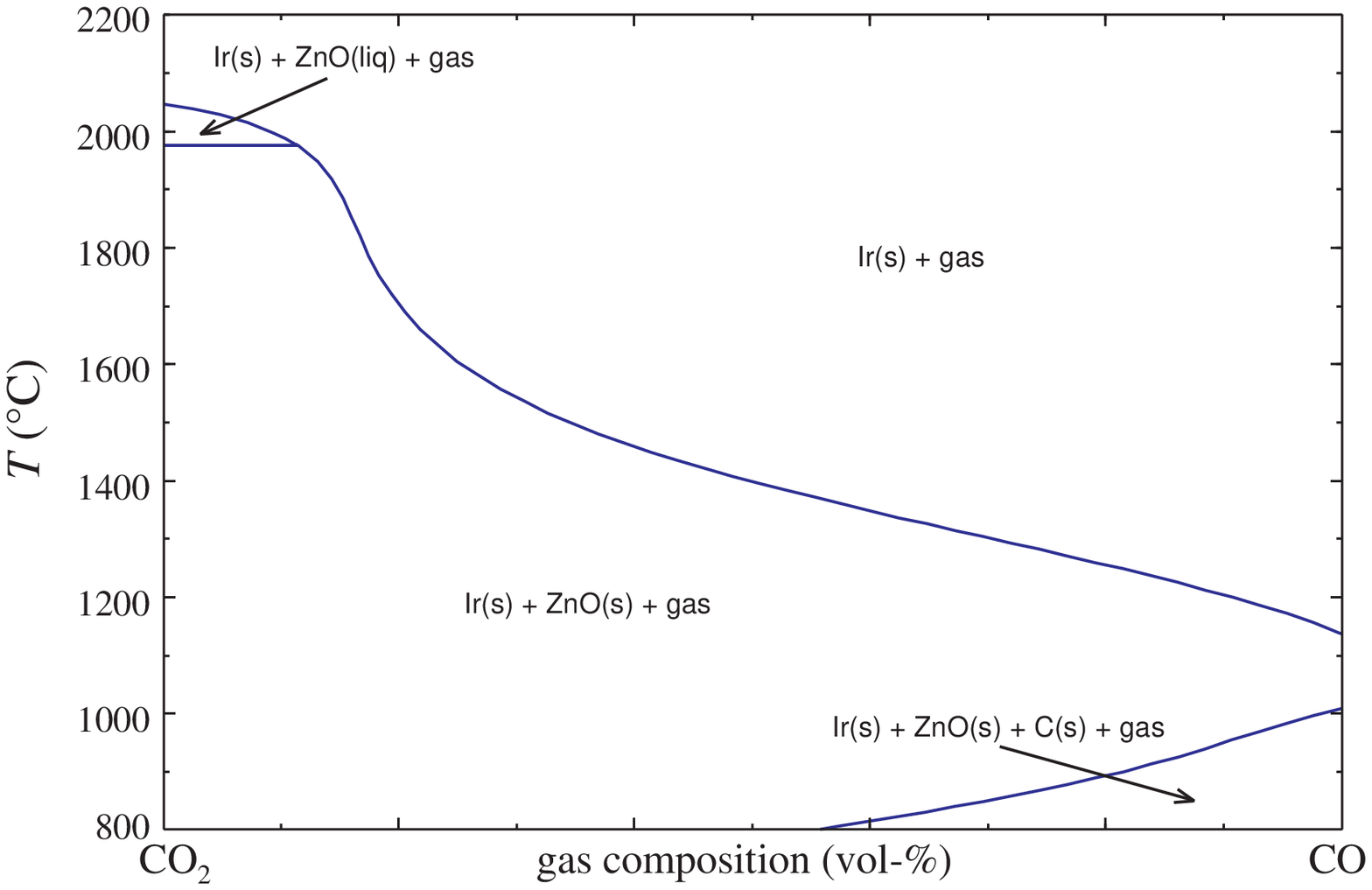} \ b)
\caption{Stability regions of phases in the system Zn -- Ir -- O -- C -- Ar in the coordinates gas composition vs. $T$ for a total pressure $p=10$\,bar. a) for gas mixtures of O$_2$ and Ar. b) for gas mixtures of CO$_2$ and CO.}
\label{fig:zn-ir-gas}
\end{figure}

\begin{description}
	\item[Gallium(III) oxide:] Tomm et al. \cite{Tomm00} reported the first successful Czochralski growth of $\beta$-Ga$_2$O$_3$. The melting point $T_\mathrm{f}=1795^{\,\circ}$C \cite{Barin93} is well beyond the stability limit of platinum, and hence Ir crucibles had to be used. Already in the proximity of the Ga$_2$O$_3$ stability limit (at $T_\mathrm{f}$: $p_{\mathrm{O}_2} \gtrsim10^{-7}$\,bar, Fig.~\ref{fig:t_vs_ox}) the evaporation, under formation of suboxides GaO and Ga$_2$O, becomes remarkable. At cooler parts Ga(I)$_2$O tends to sublimate and dissociate under formation of Ga(III)$_2$O$_3$ and Ga metal. The latter easily alloys with Ir and destroys the crucible. A sufficiently large oxygen concentration is known to suppress suboxide formation, but does unfortunately affect Ir by oxidation -- especially in colder parts of the setup (e.g. afterheater). An atmosphere of 90\% Ar + 10\% CO$_2$ supplies a $p_{\mathrm{O}_2}(T)$ that enables crystal growth without affecting Ir crucibles.
	\item[Ti:sapphire:] Under ambient conditions titani\-um forms the stable Ti$^{4+}$O$_2$, but lasing in Ti:sapphire is obtained by Ti$^{3+}$ and is suppressed by Ti$^{3+}$-Ti$^{4+}$ pairs \cite{Sanchez88}. Usually, Al$_2$O$_3$ crystals with titanium doping are grown in vacuum and the desired Ti$^{3+}$ is obtained by subsequent annealing in a reducing atmosphere (e.g. H$_2$ or CO). One should keep in mind that the term ``vacuum'' means a state that is not well defined from the thermodynamic point of view -- always some residual $p_{\mathrm{O}_2}$ will be present depending on level of vacuum and composition of the residual gas. Uecker et al. \cite{Uecker06,Bertram05} proposed an alternative technology based on Ar/CO mixtures with $p=20$\,mbar that allow the adjustment of the oxygen partial pressure $p_\mathrm{O}$ (O atoms are in excess of O$_2$ molecules under the growth conditions) to the desired range $2.0\times10^{-8}\leq p_\mathrm{O} \,\mathrm{(bar)}\leq2.3\times10^{-7}$.
	\item[Iron(II) oxide:] Depending on $p_{\mathrm{O}_2}$, $T$, and chemical environment, iron acquires usually the oxidation states Fe$^0$ (Fe metal, carbonyl Fe(CO)$_5$, carbide Fe$_3$C), Fe$^{2+}$ (wustite Fe$_{1-x}$O), Fe$^{3+}$ (hematite Fe$_2$O$_3$), Fe$^{4+}$ (strontium ferrate(IV) SrFeO$_3$ \cite{Takeda86}), or Fe$^{6+}$ (potassium ferrate(VI) K$_2$FeO$_4$ \cite{Hrostowski50}). Magnetite (iron(II,III) oxide Fe$_3$O$_4$) has a large stability field ($p_{\mathrm{O}_2},T$) under mildly oxidizing conditions, whereas the stability field of wustite is small and restricted to $p_{\mathrm{O}_2}\lesssim10^{-8}$\,bar that can hardly be maintained under the conditions of crystal growth from the melt in the technically available ``inert'' gas. Residual oxygen within Ar (99.999\% purity) is sufficient to oxidize wustite to magnetite \cite{Klimm05b}. Contrarly, the oxygen partial pressure $p_{\mathrm{O}_2}(T)$ of a gas mixture 85\% Ar, 10\% CO$_2$ + 5\% CO is ``self adjusting'' over an extended range $\mathit\Delta T$ $>1000$\,K which allows the growth of wustite ($T_\mathrm{f}=1371^{\,\circ}$C) as well as olivine (Mg,Fe$^{2+}$)SiO$_4$ ($T_\mathrm{f}\approx1890^{\,\circ}$C) crystals \cite{Ganschow05}.
	\item[Zinc oxide:] The triple point of ZnO is $1975^{\,\circ}$C, 1.06\,bar. Fig.~\ref{fig:t_vs_ox} shows that the sensibility of ZnO with respect to reduction is intermediate compared to Ga$_2$O$_3$ and Bi$_2$O$_3$. Fortunately, the $p_{\mathrm{O}_2}(T)$ supplied by CO$_2$ is sufficient to stabilize ZnO. In companion papers was reported recently on the melt growth of ZnO boules from Ir crucibles in a Bridgman-like setup \cite{Bertram04c,Schulz06,Schulz08,Klimm08}. During crystal growth $T$ gradients must be established, hence compatibility of the materials must be guaranteed over a certain $T$ range. Fig.~\ref{fig:zn-ir-gas}a shows that any fixed mixture of Ar and O$_2$ will not stabilize iridium metal Ir(s) and solid or liquid zinc oxide (ZnO(s), ZnO(liq)) for an extended $T$ range. For $T\lesssim1300^{\,\circ}$C Ir would oxidize to IrO$_2$. In contrast, Fig.~\ref{fig:zn-ir-gas}b shows that CO$_2$/CO mixtures with $\geq90$\% CO$_2$, including pure CO$_2$, can stabilize Ir together with ZnO in both condensed phases over the entire $T$ range of importance for melt growth.
\end{description}

\section{Conclusion}

It is often reported that crystal growth experiments are performed in an ``inert atmosphere''. This term is vague, as each gas that is technically available, including noble gases, contains residual impurities resulting in a small and often unknown residual oxygen partial pressure. Often the residual gas is air, resulting in $p_{\mathrm{O}_2}\approx2\times10^{-6}$\,bar. Without giving a quantitative explanation, Mateika and Rusche \cite{Mateika77b} reported that the mass loss of an empty Ir crucible heated to $1750^{\,\circ}$C was in pure CO$_2$ by ca. 40\% lower compared to an atmosphere of 98\% N$_2$ + 2\% O$_2$. A quantitative thermodynamic treatment shows that dynamic atmosphere compositions, where oxygen is produced by equilibrium chemical reactions between gases, can produce variable $p_{\mathrm{O}_2}(T)$ in such a way, that the specific oxidation state $m$ of a desired MeO$_{m/2}$ is stable over the whole $T$ range of the crystal growth process. Moreover, properly chosen atmospheres based on CO$_2$ allow the application of Ir crucibles under highly oxidative conditions. Such conditions allow the growth of oxides like ZnO which would otherwise easily be reduced to the metal.


\begin{thebibliography}{99}

\bibitem{FactSage5_5}
FactSage 5.5, \ttfamily http://www.factsage.com/ \normalfont (2007).

\bibitem{Pelton91}
A. D. Pelton in: Materials Science and Technology, Vol.~5, VCH, Weinheim, 1991 (Ed. R.~W. Cahn et al.) 1--73.

\bibitem{Klimm99c}
D. Klimm, W. Schr\"oder, J. Korean Assoc. Crystal Growth 9 (1999) 360--364.

\bibitem{Barin93}
I. Barin, Thermodynamic Data of Pure Substances, VCH, Weinheim, 1993.

\bibitem{Thermocalc08}
Thermocalc, \ttfamily http://www.thermocalc.com/ \normalfont (2008).

\bibitem{Kehl52}
W. L. Kehl, R. G. Hay, D. Wahl, J. Appl. Phys. 23 (1952) 212--215.

\bibitem{Atkins06}
P. Atkins, J. de~Paula, Atkins' Physical Chemistry, Oxford University Press 2006, 215.

\bibitem{Klimm99b}
D. Klimm, thermochimica acta 339 (1999) 111--116.

\bibitem{Sakowska01}
H. Sakowska, M. Swirkowicz, K. Mazur, T. Lukasiewicz, A. Witek, Cryst. Res. Technol. 36 (2001) 851--858.

\bibitem{Berkowski00}
M. Berkowski, J. Fink-Finowicki, W. Piekarczyk, L. Perchu\'c, P. Byszewski, L. O. Vasylechko, D. I. Savytskij, K. Mazur, J. Sass, E. Kowalska, J. Kapu\'sniak, J. Crystal Growth 209 (2000) 75--80.

\bibitem{Digiuseppe80}
M. A. {DiGiuseppe}, S. L. Doled, W. M. Wenner, J. E. Macur, J. Crystal Growth 49 (1980) 746--748.

\bibitem{Pajaczkowska99}
A. Pajaczkowska, A. Klos, D. Kasprowicz, M. Droz\-dowski, J. Crystal Growth 198/199 (1999) 440--443.

\bibitem{Pajaczkowska01}
A. Pajaczkowska, A. V. Novosselov, G. V. Zimina, J. Crystal Growth 223 (2001) 169--174.

\bibitem{Tomm00}
Y. Tomm, P. Reiche, D. Klimm, T. Fukuda, J. Crystal Growth 220 (2000) 510--514.

\bibitem{Sanchez88}
A. Sanchez, A. J. Strauss, R. L. Aggarwal, R. E. Fahey, IEEE J. Quantum Electronics 24 (1988) 995--1002.

\bibitem{Uecker06}
R. Uecker, D. Klimm, S. Ganschow, P. Reiche, R. Bertram, M. Ro{\ss}berg, R.~Fornari, SPIE Proc. 5990 (2005) 53--61.

\bibitem{Bertram05}
R. Bertram, S. Ganschow, D. Klimm, P. Reiche, R. Uecker, Patent DE 10 2005 043 398 (2005), in German.

\bibitem{Takeda86}
Y. Takeda, K. Kanno, T. Takada, O. Yamamoto, M. Takano, N. Nakayama, Y. Bando, J. Solid State Chem. 63 (1986) 237--249.

\bibitem{Hrostowski50}
H. J. Hrostowski, A. B. Scott, J. Chem. Phys. 18 (1950) 105--107.

\bibitem{Klimm05b}
D. Klimm, S. Ganschow, J. Crystal Growth 275 (2005) e849--e854.

\bibitem{Ganschow05}
S. Ganschow, D. Klimm, Cryst. Res. Technol. 40 (2005) 359--362.

\bibitem{Bertram04c}
R. Bertram, S. Ganschow, D. Klimm, P. Reiche, R. Uecker, Patent DE 10 2004 003 596, in German.

\bibitem{Schulz06}
D. Schulz, S. Ganschow, D. Klimm, M. Neubert, M. Ro{\ss}berg, M. Schmidbauer, R. Fornari, J. Crystal Growth 296 (2006) 27--30.

\bibitem{Schulz08}
D. Schulz, S. Ganschow, D. Klimm, K. Struve, J. Crystal Growth 310 (2008) 1832--1835.

\bibitem{Klimm08}
D. Klimm, S. Ganschow, D. Schulz, R. Fornari, J. Crystal Growth (2008) in press, \newline doi:10.1016/j.jcrysgro.2008.02.027.

\bibitem{Mateika77b}
D. Mateika, Ch. Rusche, J. Crystal Growth 42 (1977) 440--444.

\end{thebibliography}
\end{document}